\documentclass[11pt]{article}

\usepackage{graphicx}
\usepackage[super,compress]{cite}
\usepackage[dvips,breaklinks]{hyperref}
\usepackage{slashed}
\usepackage{bbold}
\usepackage{dsfont}

\usepackage{amsmath,amssymb,booktabs}
\hypersetup{colorlinks,urlcolor=black,citecolor=black,linkcolor=black,filecolor=black}
\usepackage{breakurl}

\renewcommand{\not}[1]{#1 \hskip-0.475em /}
\def\bib{B\kern-.05em{I}\kern-.025em{B}\kern-.08em}
\def\btex{B\kern-.05em{I}\kern-.025em{B}\kern-.08em\TeX}

\newcommand{\gsim}{\;\rlap{\lower 3.5 pt \hbox{$\mathchar \sim$}} \raise 1pt
 \hbox {$>$}\;}
\newcommand{\lsim}{\;\rlap{\lower 3.5 pt \hbox{$\mathchar \sim$}} \raise 1pt
 \hbox {$<$}\;}

\begin{document}
\thispagestyle{empty}
\renewcommand{\thefootnote}{\fnsymbol{footnote}}

\begin{flushright}
{\small
TUM-HEP-940/14\\
OUTP-14-07P\\
April 24, 2014}
\end{flushright}

\vskip1.5cm
\begin{center}
{\Large\bf\boldmath 
Muon anomalous magnetic moment and penguin loops\\[0.1cm] in
warped extra dimensions}\footnote{To appear in: Proceedings of the 
international conference on ``Flavor Physics and Mass Generation'', 
Nanyang Technological University, Singapore, 10 -- 14 February 2014. 
}
\vspace{\baselineskip}

\vspace{1cm}
{\sc M.~Beneke}, {\sc P.~Moch}\\[5mm]
${}^a${\it Physik Department T31,
Technische Universit\"at M\"unchen,\\ 
James-Franck-Stra\ss e~1, D - 85748 Garching, Germany}\\[1cm]
{\sc J.~Rohrwild}\\[5mm]
{\it Rudolf Peierls Centre for Theoretical Physics, 
University of Oxford,\\
1 Keble Road, Oxford OX1 3NP, United Kingdom
}\\[0.5cm]

\vspace*{1cm}
\textbf{Abstract}\\
\vspace{1\baselineskip}
\parbox{0.9\textwidth}{
We describe the computation of the one-loop muon anomalous magnetic 
moment and radiative penguin transitions in the minimal and custodially 
protected Randall-Sundrum model. A fully five-dimensional (5D) framework is 
employed to match the 5D theory onto the Standard Model extended by 
dimension-six operators. The additional contribution to the anomalous 
magnetic moment from the gauge-boson exchange 
contributions is 
\[
\Delta a_\mu \approx 8.8 \,(27.2)\cdot 10^{-11}\,\times (1\,\mbox{TeV}/T)^2,
\] 
where the first (second) number refers to the minimal 
(custodially-protected) model. Here $1/T$ denotes the location of the 
TeV brane in conformal coordinates, and is  related to the mass of the 
lowest gauge-boson KK excitation by $M_{\rm KK}\approx 2.35\,T$. We also 
determine the Higgs-exchange contribution, which depends on the 5D Yukawa 
structure and the precise interpretation of the localisation of the 
Higgs field near or at the TeV brane.
}
\end{center}

\newpage
\setcounter{footnote}{0}
\renewcommand{\thefootnote}{\alph{footnote}}
\setcounter{page}{1}


\newpage

\allowdisplaybreaks

\section{Introduction}
\label{sec:intro}

The Randall-Sundrum model \cite{Randall:1999ee} with bulk Standard Model 
(SM) fields and brane-localized Higgs field offers a simultaneous 
solution to the 
gauge-gravity and flavour hierarchy problems of the SM at the 
price of introducing an 
additional curved space dimension, which would manifest itself through 
a discrete spectrum of Kaluza-Klein (KK) resonances, possibly visible at the 
Large Hadron Collider (LHC). Its phenomenology has been studied in very 
much detail, but mostly at tree level. Some of the strongest constraints 
on the SM arise, however, from effects that exist only at the loop level. 
Important examples are the anomalous magnetic and electric dipole moments, 
lepton-flavour violation and quark flavour-changing neutral current 
processes related to the radiative, chirality-violating (``penguin'') 
amplitude $f_i\to f_j \gamma$. Higgs production 
and decay is a more recent addition to the list of 
loop-induced phenomena of interest. These processes are 
now also being studied in the Randall-Sundrum model (Higgs production 
and decay \cite{Casagrande:2010si,Azatov:2010pf,Goertz:2011hj,Carena:2012fk,Malm:2013jia,Frank:2013qma,Hahn:2013nza},
lepton \cite{Csaki:2010aj,Beneke:2012ie} and 
quark flavour violation \cite{Gedalia:2009ws,Delaunay:2012cz,Blanke:2012tv}).
In this article we 
describe the computation of one-loop radiative penguin transitions, and 
the muon anomalous magnetic moment in particular, in a fully five-dimensional 
(5D) framework following Refs.~\cite{Beneke:2012ie,inprep}, 
to which we refer for the more technical details.

Observables related to flavour often provide very high lower limits on 
the value of the lowest KK resonance, which can however be avoided, if the 5D 
Yukawa matrices are not generic (anarchic). For example, in a generic 
setting, the absence of non-standard CP violation in kaon mixing requires 
the first KK gluon to be heavier than 
20~TeV~\cite{Csaki:2008zd, Blanke:2008zb}, 
far beyond the reach of the LHC. The 
modification of Higgs production and signal strengths also depend on 
the unknown 5D Yukawa couplings. More model-parameter independent limits 
follow from electroweak precision observables, especially the 
$S$-parameter (since the $T$-parameter depends on whether the model is 
endowed with custodial symmetry), and of course the non-observation of direct 
production of KK resonances at the LHC. These push the KK scale into 
the multi-TeV range.

The anomalous magnetic moment is interesting in several respects. First, 
it is very precisely measured (and in some tension with the SM prediction). 
Second, it 
receives contributions from gauge-boson and Higgs exchange in the loop. 
As we discuss below, the former are insensitive to the 5D Yukawa 
couplings and fall into the category of ``electroweak precision tests''. 
The Higgs contributions, on the other hand, are model-parameter-dependent 
similar to flavour and Higgs observables. Moreover, they are sensitive to the 
scale set by the localization of the Higgs boson near the TeV brane 
in the Randall-Sundrum space-time, an issue that is also important for the 
interpretation of Higgs production.\cite{Carena:2012fk,Malm:2013jia}
 We discuss the subtleties associated 
with this issue in the framework of a 5D calculation 
in the RS theory with unbroken electroweak symmetry. 
Last but not least, the anomalous magnetic moment in the RS model 
is a non-trivial but 
instructive case for setting up and performing loop calculations 
in a quantum field theory in curved space-time with boundaries.

\section{From the 5D bulk to a 4D effective Lagrangian}
\label{sec:effLagr}

The RS space-time consists of a slice of Anti-de-Sitter space limited 
by four-dimensional flat branes. In conformal coordinates the metric 
of the 5D bulk is 
\begin{equation}
ds^2 = \left(\frac{1}{k z}\right)^2\left(\eta_{\mu\nu} dx^\mu dx^\nu-
dz^2\right), 
\label{metric}
\end{equation}
where $k\sim M_{\rm Pl} \sim 10^{19}\,$GeV is of order of the Planck 
scale $M_{\rm Pl}$, while the four-dimensional boundaries are located at 
$z=1/k$ and $z=1/T$ with $T$ of order TeV. The explicit appearance of a 
TeV scale parameter is a coordinate artifact. The proper distance between 
the two branes, $1/k\times\ln(k/T)$, is naturally only a few 
times the Planck length when $T$ varies over many orders of magnitude.

Since the present non-observation of KK resonances requires the scale $T$  
to be much larger than the scale of the SM, set by the 
Higgs vacuum expectation value $v$, the only dynamical degrees of freedom 
below the scale $T$ are the KK zero modes, which are associated with the 
usual SM fields. We can therefore match the RS theory 
onto an effective four-dimensional theory at the scale $\mu$ 
($T\gg\mu\gg M_{\rm EW}$), whose Lagrangian consists of the SM Lagrangian 
plus SU(3)$\times$SU(2)$\times$U(1) invariant higher-dimension 
operators built from SM fields:
\begin{equation}
{\cal L}_{\rm RS}^{\rm (5D)} \quad\longrightarrow \quad 
{\cal L}_{\rm eff} = {\cal L}_{\rm SM} + \frac{1}{T^2} 
\sum_i c_i {\cal O}_i.
\label{dim6}
\end{equation}
The dominant effects are captured by dimension-six 
operators \cite{Buchmuller:1985jz,Grzadkowski:2010es}. 
Since the matching coefficients $c_i$ are dominated by 
distances $\lsim 1/T$, the Higgs bilinear term in $V(\Phi)$ can be 
treated as a perturbation, and the $c_i$ can be computed in the 
theory with unbroken electroweak gauge symmetry. This results in 
a great technical simplification for the 5D propagators of 
the gauge and fermion fields.

We note that the 5D theory is non-renormalizable 
and must itself be defined as an effective theory below a scale 
$\Lambda$ that should be at least a few times the Planck scale. It 
is generally assumed (and required to solve the hierarchy problem) 
that in the mixed representation the 
four-dimensional loop momenta should be cut-off at a value 
$\Lambda(z)$ that depends on the position $z$ in the fifth dimension. 
If $\Lambda(1/k)$ is a few times the Planck scale, then the cut-off 
$\Lambda(1/T)$ relevant to processes dominated by physics near the 
TeV brane should be a few times the TeV scale. This appears to be in  
conflict with the 5D formalism, which encodes 
the sum over all KK states rather than including only the few 
below the cut-off $\Lambda(1/T)$. However, for a finite quantity 
such as the anomalous magnetic moment of the muon, the KK sum 
must converge, and the effect of including the entire tower 
relative to the truncation is of order $T^2/\Lambda(1/T)^2$, which 
is the generic size of corrections expected from the UV completion 
of the RS model. Besides, the general framework of renormalization 
in curved space-time should apply to the RS theory treated as a 
5D quantum field theory, and makes no reference to the KK decomposition.

The matching strategy outlined above does not depend on the details of 
the 5D Lagrangian though the matching coefficients $c_i$ do. We will use 
the simplest set-up of the Randall-Sundrum model. All SM fields are 
allowed to propagate throughout the 
five-dimensional bulk, except for the Higgs doublet which is confined 
to the IR brane at $z=1/T$. No further field content is added.  
Quarks as well as the strong 
sector are not relevant for the following discussion of leptonic 
transitions.\footnote{The Lagrangian is specified in detail in 
Ref.~\cite{Beneke:2012ie}.} 
This ``minimal'' model is no longer attractive from 
the phenomenological point of view, since tree-level custodial-symmetry 
violation requires the scale $T$ to be larger than roughly 4~TeV 
\cite{Csaki:2002gy, Agashe:2003zs, Casagrande:2008hr}. It is, however, useful to illustrate 
the general approach. At the end of this article, we discuss the 
extension of the computation to the RS model with custodial symmetry.

Only a few dimension-six operators from the general expression (\ref{dim6}) 
are relevant to the leptonic radiative transitions at the one-loop 
level:
\begin{eqnarray} 
\sum_i c_i {\cal O}_i & =&
a_{B,ij} \,\bar L_i\Phi \sigma_{\mu\nu} E_j B^{\mu\nu} + 
a_{W,ij} \,\bar L_i\tau^a \Phi \sigma_{\mu\nu} E_j \,W^{a,\mu\nu} + 
\mbox{h.c.}
\nonumber\\[-0.2cm]
&&+\,
b_{ij} \,(\bar L_i\gamma^\mu L_i)(\bar E_j\gamma_\mu E_j) 
+ c_{1,i} \,(\bar E_i\gamma_\mu E_i) (\Phi^\dagger iD^\mu \Phi)
\nonumber\\[0.1cm]
&& + \,c_{2,i} \,(\bar L_i\gamma_\mu L_i) (\Phi^\dagger iD^\mu \Phi)
+ c_{3,i} \,(\bar L_i\gamma^\mu \tau^a L_i) (
\Phi^\dagger \overleftrightarrow{i \tau^a D_\mu} \Phi)
\nonumber\\[0.15cm]
&& + h^{ij}  \Phi^\dagger \Phi\; \bar L_i \Phi E_j   + \text{h.c.}\,,
\label{dimsixlagragian}
\end{eqnarray}
where $\overleftrightarrow{i \tau^a D_\mu}=1/2 \, 
(i \tau^a D_\mu - i \overleftarrow D_\mu \tau^a)$.
$L_i$ ($E_i$) represents a lepton doublet (singlet) field of flavour $i$. 
The SM Higgs doublet is denoted by $\Phi$, and $ B_{\mu\nu}$ and $W^A_{\mu\nu}$
are the field strength tensors of  U(1)$_{\textrm{Y}}$ and  
SU(2)$_{\textrm{L} }$ gauge field, 
respectively. 
 
\begin{figure} 
\begin{center} 
\includegraphics[width=0.8 \textwidth]{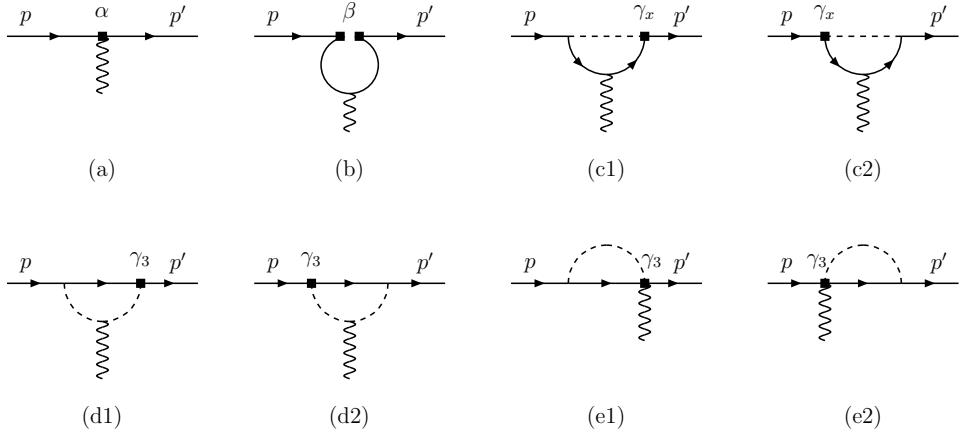}  
\end{center} 
\caption{Diagrams that contribute to $a_\mu$ in the 4D effective theory.}  
\label{fig:4Ddiags}  
\end{figure}

The last three lines of \eqref{dimsixlagragian} contain operators that
contribute to radiative penguin transitions at the one loop-level 
in the effective theory, but can be generated at tree level in 
RS model. The operators in the first line correspond to dipole 
operators after electroweak symmetry breaking and therefore 
contribute at tree level. However, they are generated only by loops 
in the 5D theory. Once the corresponding matching coefficients 
have been computed in the RS model at tree- and one-loop level, 
respectively, the amplitude for the radiative dipole transition 
is obtained by the following steps. First, we parametrize the fields 
by the mass eigenstates in the broken theory below the electroweak 
scale, making the replacement
\begin{align}\label{eq:EWSBsub}
\Phi\to \frac{1}{\sqrt{2}}\left(
\begin{array}{c} \sqrt{2}\phi^+ \\ v+H+iG \end{array}\right)&&
E_i \to V_{ij} P_R \psi_j,\qquad
L_i \to U_{ij}  P_L  \!\left(
\begin{array}{c} \nu_j \\ \psi_j \end{array}\right),
\end{align}
where $\psi_i$ is the Dirac spinor field for the massive 
leptons ($i=1,2,3$ corresponding to electron, muon, tau)
and $\nu_i$ is the corresponding neutrino spinor field. 
$P_{L/R} = (1\mp\gamma_5)/2$ are the chiral projectors.
The unitary matrices $U$ and $V$ relate the interaction and mass 
flavour bases. Then we compute the dipole transition with the 4D 
effective theory. This requires the computation of the $(p+p^\prime)^\mu$ 
structure (related to $\sigma^{\mu\nu}q_\nu$ via the Gordon identity) 
of the diagrams 
shown in Fig.~\ref{fig:4Ddiags}. The couplings in the mass basis 
are now given by
\begin{eqnarray}
\alpha_{ij} &=& [U^\dagger a V]_{ij},
\nonumber\\[0.1cm]
\beta_{ijkl} &=& \sum_{m,n} \,
[U^\dagger]_{im} U_{mj} [V^\dagger]_{kn} V_{nl} \,b_{mn} ,
\nonumber\\
\gamma_{1,ij} &=& \sum_m \,[V^\dagger]_{im} V_{mj} \,c_{1,m} ,
\nonumber\\
\gamma_{x,ij} &=& \sum_m \,[U^\dagger]_{im} U_{mj} \,c_{x,m} \quad
(x=2,3)\,,
\label{couplingdefs}
\end{eqnarray}
with $a_{ij}=c_W a_{B,ij} - s_W a_{W,ij}$.

The calculation is straightforward and can be carried out with standard
techniques. Let us mention a few subtleties:
\begin{itemize}
\item Though finite the diagrams must be evaluated in some 
 regularization scheme. In dimensional regularization terms of the form 
 $\epsilon \times \frac{1}{\epsilon}$ arise that would be missed in a 
 purely four-dimensional calculation.
\item The sum of all the one-loop diagrams in the effective theory is
 scheme dependent. This dependence cancels with
 the scheme dependence of the matching coefficient $\alpha_{ij}$ 
 of the dipole operators that arises in the 5D loop calculation.
\item In naive dimensional regularization (NDR, anti-commuting $\gamma_5$)
 only the dipole and four-lepton operators give a non-vanishing 
 contribution (diagrams (a), (b) in Fig.~\ref{fig:4Ddiags}).  
\end{itemize}

Focusing on the muon anomalous magnetic moment, 
the result in NDR scheme is given by\footnote{
The formula omits a small contribution of approximately 
$0.4\cdot 10^{-11} \times \left(1\,\mbox{TeV}/T\right)^2$ from 
the $\gamma_{x,22}$ couplings that can be interpreted as modifications of  
the gauge-boson-fermion vertices in the SM diagrams.}  
\begin{eqnarray}
\Delta a_\mu &=&  \frac{g_\mu-2}{2} = -\,\frac{4 m_\mu^2}{T^2}\left(
\frac{\mbox{Re}\,(\alpha_{22})}{y_\mu e}
+\sum_{k=1,2,3} \frac{1}{16\pi^2}\,
\frac{m_{\ell_k}}{m_\mu}\,
\mbox{Re}\,(\beta_{2kk2})
\right)\,,
\label{amures}
\end{eqnarray}
where $\Delta a_\mu$ refers to the additional contributions generated by the 
KK excitations. Other observables such as the $\mu \to e \gamma$ or 
$\tau \to e\gamma$ branching fractions can be determined in an 
analogous fashion. The electric dipole moments are related to the 
imaginary parts of the matching coefficients. Note that we extracted 
the square of the muon mass by dividing by the small 4D muon Yukawa 
coupling $y_\mu$. This anticipates that to a very good approximation 
we shall find that the gauge-boson exchange contribution to 
$\alpha_{22}$ is proportional to $y_\mu$. The natural 
size of $\Delta a_\mu$ in the RS model is therefore of order 
$\alpha_{\rm em}/(4 \pi) \times m_\mu^2/T^2 
\approx 0.7\cdot 10^{-11} \times \left(1\,\mbox{TeV}/T\right)^2$, 
far too small (unless $T$ is unrealistically small)  
to explain the present discrepancy between measurement and 
theoretical prediction, 
$ a_\mu^{\rm exp}-a^{\rm SM}_\mu=239(63)(48)\times 10^{-11}$\cite{pdg2012}, 
unless there is some additional parametric or numerical enhancement.

\section{5D Feynman rules}
\label{sec:matching}

The next and most important step is the determination of the Wilson 
coefficients in \eqref{dimsixlagragian} from the underlying RS model. 
We perform the matching calculation in a manifestly 5D formalism rather 
than work with  an infinite tower of KK modes. This avoids the calculation 
of KK sums but requires the knowledge of the 5D Feynman rules.

The general strategy for their derivation was established 
in Refs.~\cite{Randall:2001gb,Csaki:2010aj}. We refer the reader to the 
appendix of Ref.~\cite{Beneke:2012ie} for a comprehensive summary. Since the 
RS model has ordinary 
translation invariance on four-dimensional hypersurfaces 
orthogonal to the fifth dimension, 
it is convenient to use a mixed momentum-coordinate
space representation with four-dimensional, continuous momentum and 
a bulk position variable. The necessary ingredients for the calculation 
are then: 5D vertex rules, 5D propagators and 
the zero-mode wave functions. Since we integrate out scales far above the 
electroweak scale and match onto a set of SU(3)$\times$SU(2)$\times$U(1) 
invariant operators, it suggests itself to work in the 
SU(2)$\times$U(1)-symmetric phase. The wrong-sign Higgs mass term 
is then a perturbation, and the zero-mode fermions and gauge bosons 
do not receive masses due to the Higgs mechanism. Their bulk 
wave functions  
are simply given by~\cite{Pomarol:1999ad,Grossman:1999ra}
\begin{align}
 f_\gamma^{(0)}(z) = \sqrt{\frac{k}{\ln\frac{k}{T}}}\,, \quad &&
f_L^{(0)}(z) = 
\sqrt{\frac{1-2c_L}{1-(\frac{T}{k})^{1-2c_L}}}\sqrt{T}\,(kz)^2(Tz)^{-c_L}
\end{align}
and a similar expression with $c_{L_i} \to - c_{E_i}$ for the right-handed 
zero-mode $g_{E_i}(z)$ of the singlet lepton field $E_i$.\footnote{
$c_\psi = M_\psi/k$ denotes the dimensionless parameter related to the 
Planck-scale 5D bulk mass $M_\psi$ of fermion field $\psi$. In general, 
for fermions, $f$ ($g$) denotes left-handed (right-handed) mode functions 
from the 4D perspective.} Note 
that reference to the KK zero modes cannot be avoided, 
since they correspond to the SM fields, which are not integrated out 
(unless highly virtual). Their wave functions appear on the external legs 
when the 5D Green functions are matched to those of the 4D effective 
theory.

Determining the Feynman rules is in principle straightforward. While the 
vertices are simple, the propagators require more work. As usual, they 
are found by inverting the differential operator in the bilinear terms 
of the action. For fermions, in the mixed representation, this requires 
solving
\begin{equation}
\left[\frac{1}{kz}\right]^4\mathcal{D}\Delta(p,z,z')=i\delta(z-z')
\mathds{1}\qquad \mbox{with} \qquad
\mathcal{D}= \slashed p + 
i\Gamma^5(\partial_z-\frac{2}{z})-\frac{c}{z}\,.
\end{equation}
Since the 5D fermions are non-chiral, the propagator contains four chiral 
components, 
\begin{eqnarray}
\Delta_L(p,x,y) &=&  
-\underbrace{P_L F^+_L(p,x,y) \slashed{p} P_R}_{\rm contains\;zero\;mode}
-P_R F^-_L(p,x,y) \slashed{p} P_L
\nonumber \\
&& +\,\underbrace{ P_L d^+F^-_L(p,x,y)P_L + P_R d^-F^+_L(p,x,y)P_R}_{
\rm mass \;terms}\,,
\label{chiralfermionprop}
\end{eqnarray}
where 
\begin{equation}
\left[-p^2-\partial_z^2+\frac{c^2\pm c-6}{z^2}
   +\frac{4}{z}\partial_z\right] F^\pm(p,z,z') = i(kz)^4 \delta(z-z')
\label{DEprop}
\end{equation} 
and $d^\pm \equiv \pm [\partial_z-(2\pm c)/z]$.
The interpretation is clear when written in terms of the KK decomposition, 
which reads, for example: 
\begin{eqnarray}
&& F^+_{L}(p,x,y) =\sum_{n}
\,f^{(n)}_L(x)\frac{-i}{p^2-m_n^2}\,f^{(n)}_L(y) 
\nonumber \\ 
&& d^-F^+_{L}(p,x,y) =\sum_{n}
\,g^{(n)}_L(x)\frac{im_n}{p^2-m_n^2}\,f^{(n)}_L(y) 
\end{eqnarray}
In the 5D formalism, we solve (\ref{DEprop}) in terms of Bessel functions. 
Introducing 
\begin{eqnarray}
S_\pm(p,x, y,  c) &=& I_{c \pm 1/2}(px) K_{c \pm 1/2}(py) - 
                 K_{c \pm 1/2}(px) I_{c \pm 1/2}(py)
\nonumber \\[0.1cm]
\tilde{S}_\pm(p,x, y,  c) &=& I_{c \pm 1/2}(px) K_{c \mp 1/2}(py) + 
                 K_{c \pm 1/2}(px) I_{c \mp 1/2}(py)\,,
\end{eqnarray}
the Euclidean propagator is given by expressions such as 
\begin{eqnarray}
d^+F^-_L(p, x, y) &=& 
- p \,\Theta(x - y)\,\frac{i  k^4 x^{5/2} y^{5/2} 
\tilde{S}_+(p, x,1/T, c_L)S_-(p,y, 1/k,c_L)}
{S_-(p,1/T, 1/k , c_L)} \nonumber \\
&& - \,p \,\Theta(y - x) \,
\frac{i k^4 x^{5/2} y^{5/2} S_-(p,y, 1/T,c_L) \tilde{S}_+(p, x, 1/k , c_L)}
{S_-(p,1/T, 1/k , c_L)}\,.\qquad
\end{eqnarray}
We note (for later) that the propagator is discontinuous at $x=y$.
Similar expressions are found for the other propagator components and 
for SU(2) singlet fermion and 
gauge fields.

\section{Matching: Tree-level operators}
\label{sec:treeoperators}

\begin{figure} 
\begin{center} 
\vskip0.3cm
\includegraphics[width=3cm]{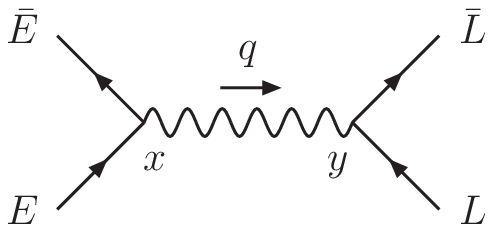}\hskip1cm   
\includegraphics[width=3cm]{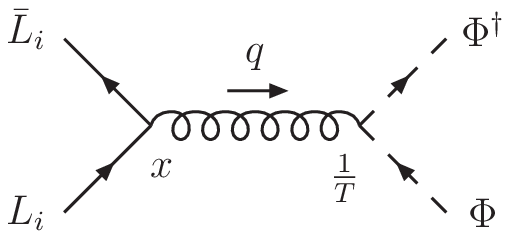}  
\end{center} 
\caption{Left: Hypercharge boson exchange that generates the 
four-fermion operator $(\bar L_i\gamma^\mu L_i)(\bar E_j\gamma_\mu E_j)$. 
Right: SU(2) gauge boson exchange that generates the 
fermion-Higgs operators in (\ref{dimsixlagragian}).}  
\label{fig:treetopology}  
\end{figure}

With the 5D propagators at hand the determination of the tree-level 
matching coefficients of the  
four-lepton operator $(\bar L_i\gamma^\mu L_i)(\bar E_j\gamma_\mu E_j)$ 
and the lepton-Higgs operators in (\ref{dimsixlagragian}) becomes 
trivial. The relevant diagrams are shown in Fig.~\ref{fig:treetopology}.

Since the lepton-Higgs operators do not contribute to the anomalous 
magnetic moment in the NDR scheme, see (\ref{amures}), we only discuss 
the four-lepton operator. It can be generated only by hypercharge gauge 
boson exchange. The diagram readily translates into an expression for the 
Wilson coefficient:  
\begin{eqnarray}\label{4fermioncoeff}
b_{ij} &=& -i \,(i g_5^{\prime})^2 \,\frac{Y_L}{2}\frac{Y_E}{2} \,
T^2 \int_{1/k}^{1/T} \!\!dx dy \,
\frac{{f_{L_i}^{(0)}}^2\!(x)}{(k x)^4}\,
\frac{{g_{E_j}^{(0)}}^2\!(y)}{(k y)^4}\,
\Delta_\perp^{\rm ZMS}(q=0,x,y)\;. 
\end{eqnarray}
$\Delta_\perp^{\rm ZMS}$ refers to the 5D propagator with the 
massless zero-mode subtracted, since zero-mode exchange is a low-energy 
effect and not part of the matching coefficient. Once the zero mode is 
subtracted, the external momenta can be set to zero, implying $q=0$. 
In this limit, the gauge-boson propagator $\Delta_\perp^{}(q,x,y)$ has a 
particularly simple structure:
\begin{eqnarray}
\Delta_\perp(q,x,y) &\stackrel{q\to 0}{=}& 
\Theta(x-y) \,\frac{i k}{\ln\frac{k}{T}} 
\,\bigg(-\frac{1}{q^2} + \frac{1}{4}\,
\bigg\{\frac{1/T^2-1/k^2}{\ln\frac{k}{T}} - x^2-y^2
+ 2 x^2 \ln(x T)\nonumber\\
&& \hspace*{0cm} + \, 
2 y^2 \ln(y T)+2 y^2 \ln\frac{k}{T}
\bigg\} + {\cal O}(q^2) \bigg) 
+ (x\leftrightarrow y).
\label{gaugepropZeroMom}
\end{eqnarray} 
The singular piece $\propto 1/q^2$ arises from the massless zero mode 
that needs to be subtracted. After inserting \eqref{gaugepropZeroMom} 
into \eqref{4fermioncoeff} all integrals are elementary.
The result agrees with the calculation of four-quark operators in the 
KK mode language \cite{Casagrande:2008hr}.

Inserting the Wilson coefficient into \eqref{amures} gives a contribution 
to the anomalous magnetic moment of   
\begin{equation}\label{fourfermioncontribution}
\Delta a_\mu  = \frac{\alpha_{\rm em}}{8\pi c_W^2}  
\,\frac{m_\mu^2}{T^2} \,\frac{1}{\ln\frac{k}{T}}\,f(\ln(k/T),c_L,c_E)
\approx 1.2 \times 10^{-13}\times 
\frac{\left(1\,{\rm TeV}\right)^2}{T^2}\;.
\end{equation}
The function $f$ is close to one, resulting in the numerical 
estimate above. This is far below the current experimental and 
theoretical uncertainties for any allowed value of the KK scale 
$T$, since it is suppressed relative to the naive estimate by the 
large factor $\ln\frac{k}{T} \approx 35$. 

\begin{figure}[t] 
\begin{center} 
\vskip-0.0cm
\includegraphics[width=10.5cm]{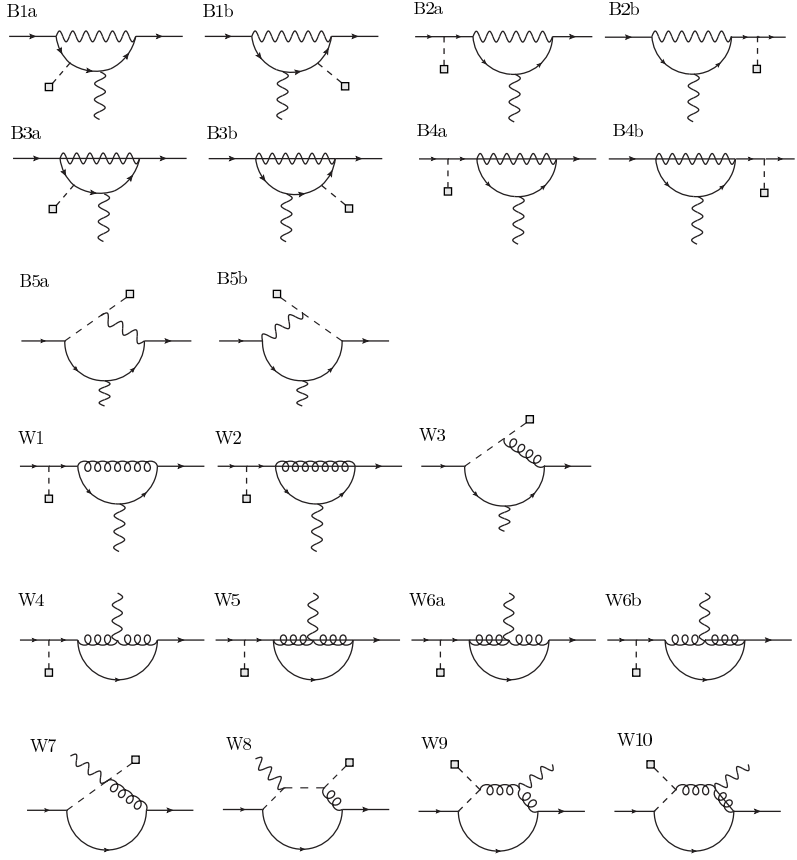}  
\end{center} 
\caption{Diagrams contributing to the matching
  coefficients of the dimension-six dipole operators. Solid lines refer to leptons,
with the right external line belonging to the doublet $L_i$, 
the left one to $E_j$. Wavy lines denote hypercharge gauge bosons and 
the external photon, curly lines SU(2) W-bosons. A solid-wavy 
(solid-curly) line refers to the scalar fifth component  
of the gauge field. Dashed lines denote Higgs bosons, including the 
external Higgs field (grey box). Vertices involving Higgs fields 
are localized at $1/T$, all other vertices are integrated over 
position in the fifth dimension.}  
\label{fig:diagsall}  
\end{figure}  

\section{5D penguin diagrams -- gauge boson exchange}

The RS model does not generate the dimension-six dipole
operators at tree-level. The matching calculation for the dipole 
operators therefore requires the calculation of genuine 5D one-loop 
diagrams. There are two classes of diagrams that enter the matching 
procedure: diagrams with an internal gauge-boson propagator, shown in 
Fig.~\ref{fig:diagsall}, and diagrams with internal Higgs 
exchange. We first discuss the gauge-boson contributions, which 
are technically more difficult but conceptually simpler than the 
Higgs diagrams, discussed in the following section.

\subsection{The calculation}
\label{sec:calculation}

The calculation of the one-loop coefficients $a_{ij}$ can be simplified 
by restriction to an external photon, i.e. the
linear combination $c_W B_\mu + s_W W^3_\mu$, which 
reduces the number of diagrams (slightly). In addition, we ignore 
from the start terms that vanish when the 
Higgs doublet in the operators $\bar L_i\Phi \sigma_{\mu\nu} E_j B^{\mu\nu}$, 
$\bar L_i\tau^a \Phi \sigma_{\mu\nu} E_j \,W^{a,\mu\nu}$ is set to 
its vacuum expectation value. All non-vanishing  one-loop diagrams are 
shown in Fig.~\ref{fig:diagsall}. 

The matching coefficients must only absorb quantum effects 
related to the short distance scales $T$ and $k$. In general, however, 
a one-loop diagram has three distinct parts:
\begin{itemize}
\item[(1)] A part where each propagator only propagates the zero mode. 
  This is obviously part of the SM contribution to the penguin amplitude 
  and must be removed. It turns out that this can be achieved by subtracting 
  the zero mode from only the gauge-boson propagators. The reason is 
  that the presence of a gauge-boson zero mode automatically forces all 
  other propagators to only propagate a zero-mode due to orthogonality 
  relations and the fact that all external states are zero modes. 
\item[(2)] At least one propagator contains a KK mode, but the 4D loop 
  momentum $l$ is much smaller than the the scale $T$. The subgraph 
  consisting of KK mode propagators can be contracted to point, which 
  corresponds to the insertion of a higher-dimensional operator into a 
  4D graph. In most cases (as when more that one KK propagator is present)
  the corresponding operator is of a dimension higher than six and can 
  be ignored. The remaining dimension-six operator insertions precisely 
  correspond to the one-loop matrix elements of the non-dipole 
  operators in (\ref{dimsixlagragian}) with tree-level matching coefficients 
  as determined in the previous section. An example is diagram B1a/b in 
  Fig.~\ref{fig:diagsall}. The gauge-boson propagator must propagate 
  KK modes (see above). If the fermion lines are zero modes then the 
  short-distance subgraph is the four-lepton operator, and the contracted 
  diagram corresponds to the insertion of this operator as in diagram (b) 
  of Fig.~\ref{fig:4Ddiags}.
\item[(3)] Finally, we have the contribution where the loop momentum is 
 of the order $l\sim T$ or larger.
 Only this is part of the Wilson coefficient $a_{ij}$.   
 This contribution can be extracted directly by expanding the integrand 
 in the lepton external momenta $p$ and $p^\prime$ (after subtracting 
the zero mode from the gauge-boson propagator). 
 The expansion is usually only necessary to the first non-trivial order 
 as a dipole operator  
 $\bar L \Phi \sigma^{\mu\nu} E \, F_{\mu\nu}$ is linear in the external 
 momenta (see, however, below).
\end{itemize}        

The remaining calculation is tedious and requires a combination of further 
analytical simplifications and final numerical integrations. To give an 
example, we consider again diagram B1. We decompose each fermion propagator into its four chiral components using (\ref{chiralfermionprop}). Most of the 
64 possible terms vanish due to the chiral projectors $P_R$ 
and the brane boundary 
conditions $g_{L_i}(1/T)=f_{E_j}(1/T)=0$. 
The two remaining terms result in 
\begin{eqnarray}
  {\rm {\bf B1a}} &=& \frac{g'^2_5 e_5 Q_\mu Y_{L} Y_{E}
    y^{\rm(5D)}_{{ij}} T^3}{4 k^3} \int^{1/T}_{1/k}
  \!\!\!\frac{dx}{(kx)^4} \int^{1/T}_{1/k}
  \!\!\!\frac{dy}{(ky)^4} \int^{1/T}_{1/k}
  \!\!\!\frac{dz}{(kz)^4}\int
  \!\!\! \frac{d^4l}{(2\pi)^4}  \nonumber \\
  &&  f^{(0)}_{L_i}(z) 
  f^{(0)}_\gamma(y) g^{(0)}_{E_j}(x) \epsilon^{*\mu}  \Delta^{\rho\nu}_{\mbox{\tiny ZMS}}(l,x,z)
  \nonumber \\
  && \bar{L_{i}}(p^\prime) \Big[ F^+_{L_i}({\hat{p}^{\,\prime}},z,y) F^+_{L_i}({\hat{p}},y,1/T) F^-_{E_j}({\hat{p}},1/T,x) \left\{ \gamma_\rho (\not\!p^{\,\prime}-\!\!\!\not l) \gamma_\mu \gamma_\nu\right\}(p-l)^2 + \nonumber \\  
  && d^+F^-_{L_i}({\hat{p}^{\,\prime}},z,y) d^-F^+_{L_i}({\hat{p}},y,1/T) F^-_{E_j}({\hat{p}},1/T,x) 
  \left\{ \gamma_\rho \gamma_\mu (\not\!p-\!\!\!\not l) \gamma_\nu
  \right\} \Big] P_{R} E_{{j}}(p),\qquad
\label{diag1text2}
\end{eqnarray}
where $\hat{p} = p-l$, $\hat{p}^{\,\prime} = p^\prime-l$. This provides 
the starting point for the above-mentioned expansion in $p$ and $p^\prime$. 
The numerical integrations include the modulus of the 4D Euclidean loop 
momentum $l$ and the three bulk coordinates. Considerable numerical speed-up 
and improved accuracy can be achieved by carrying out the integration 
of the photon vertex bulk position $y$ analytically. We refer to 
Ref.~\cite{Beneke:2012ie} for the details and add here only a few comments.

{\em Scheme (in)dependence} -- 
The scheme dependence of the one-loop diagrams (b) to (e) in 
Fig.~\ref{fig:4Ddiags} must be cancelled by the scheme dependence of 
the one-loop dipole coefficient $a_{ij}$. The scheme-dependence of the 
short-distance contributions arises from diagrams B1a/b and W8 
that are potentially IR singular {\it after} the expansion in the external 
momenta $p$ and $p^\prime$.  As was the case for diagrams (b) to (e), there 
is in fact no singularity due to evanescent numerators. However, a 
naive treatment in $d=4$ dimension misses finite terms of the form 
$\varepsilon_{IR} \times \frac{1}{\varepsilon_{\rm IR}}$. These terms have 
to be added ``by hand'' by computing analytically the difference between 
the correct $d$-dimensional and naive four-dimensional treatment. 
They precisely 
cancel the scheme dependence due to the tree-level operator insertions.
After this we can safely work in $d=4$ dimensions
and use numerical methods to determine the short-distance coefficients.     

{\em One-particle reducible (1PR), ``off-shell'' contributions} -- 
1PR diagrams such as B2a contribute to the short-distance 
coefficient. The fermion propagator that connects the external Higgs 
insertion to the loop is 
\begin{eqnarray}
\Delta_i^L(p,x,1/T)P_R &=&-F_{L_i}^+(p,x,1/T) \slashed{p} P_R
+ d^-F_{L_i}^+(p,x,1/T) P_R\,.
\end{eqnarray}
The second term on the right-hand side propagates only KK excitations 
and is purely short-distance. The first term vanishes by the 
on-shell condition $\slashed{p} u(p,s)=0$ except for the zero-mode 
contribution 
\begin{align}
 \Delta_{\rm ZM}(p,x,1/T)=f^{(0)}_{L_i}(x)
\frac{i \slashed{p}}{p^2}P_R f^{(0)}_{L_i}(1/T)\,.
\end{align}
If the one-particle pole at $p^2=0$ remains in the final answer, this is a 
clear sign for a long-distance effect and this part of the external 
Higgs insertion into a zero mode needs to be amputated. (In fact there is 
an infinite series of external Higgs insertions connected by zero-mode 
propagators that necessarily sums to 
the massive SM lepton propagator.) However, when the  $1/p^2$ factor 
is cancelled by numerators, we obtain an additional contribution to the 
matching coefficient. If we parametrize the one-particle irreducible 
$\bar L_i L_i\gamma$ vertex subdiagram with off-shell zero-mode fermions 
as 
\begin{equation}
\label{OffShellVertex}
\Lambda^\mu = \Lambda^\mu_{\rm on}
                   + \slashed{p}' \Lambda^\mu_{\rm off,\,p'} 
                   + \Lambda^\mu_{\rm off,\,p} \,\slashed{p}\,,
\end{equation} 
the piece of interest arises from the 
$\Lambda^\mu_{\rm off,\,p} \,\slashed{p}$ term, and is 
given by 
\begin{eqnarray}
&& \hspace*{-0.3cm}\Lambda^\mu \Delta_{\rm ZM}(p,x,1/T)
= i\,\Lambda^\mu_{\rm off,\,p} \, 
f_{L_i}^{(0)}(x) f_{L_i}^{(0)}(1/T)\,.
\end{eqnarray}
Note that to compute this piece for the dipole transition, we need 
to expand the diagram to second order in the external momenta to 
extract the coefficient of the $p^\mu$ and $p^{\prime\mu}$ terms. 
For the standard choices of the bulk mass parameters of the 
left- and right-handed fermions, we find that these ``off-shell'' terms 
are numerically suppressed. 

\begin{figure} 
\begin{center} 
\vskip-0.1cm
\includegraphics[width=0.4\textwidth]{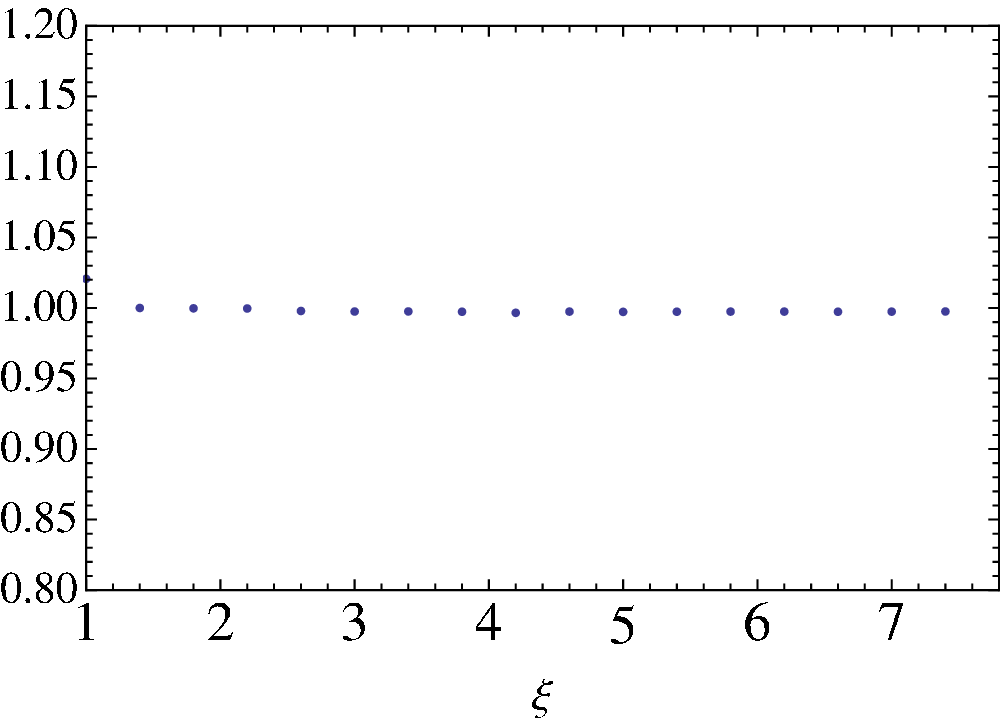}  
\hspace*{0.4cm}
\includegraphics[width=0.4\textwidth]{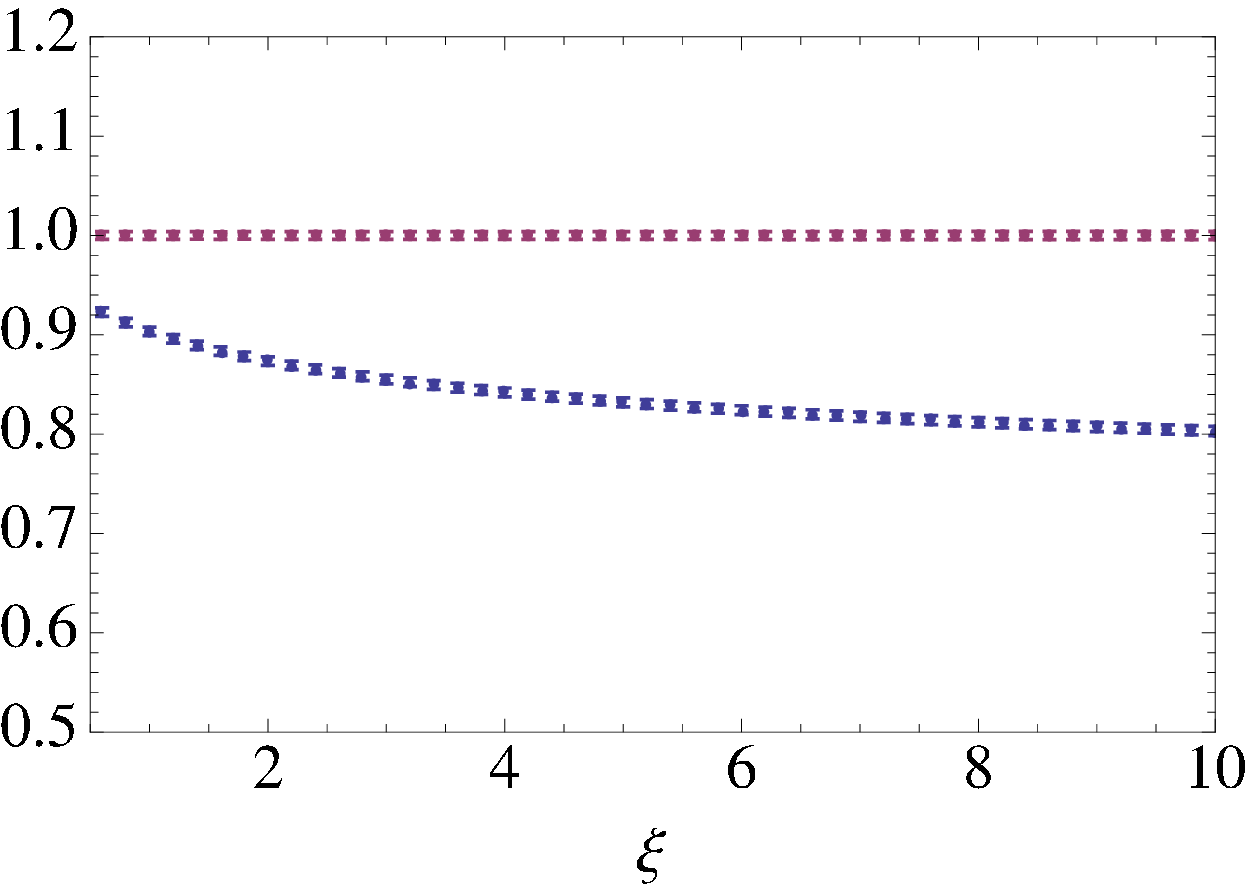}  
\end{center} 
\caption{{\it Left panel:} 
Residual gauge-parameter dependence of the short-distance coefficient
$a_{22}$ in the minimal RS model normalized to its 
value for $\xi=1$ for $c_{L_2}=-c_{E_2}=0.5748$. 
{\it Right panel:} As left panel but for the gauge-invariant 
subset of all $Z_X$-diagrams in the 
custodially protected RS model for $c_{L_i}=0.1$ and $c_{E_j}=-1.2$. 
The lower curve is without off-shell
terms, the upper line includes the off-shell terms.}  
\label{fig:gaugedep}  
\end{figure}  

{\em Gauge invariance} -- 
We performed the computation in general covariant 5D gauge and 
verified the gauge-parameter independence analytically by using 
algebraic equation-of-motion and integration-by-parts identities. 
A sketch of the proof is given in Ref.~\cite{Beneke:2012ie}. 
The 1PR contributions are required to make the 
result gauge invariant. We also 
perform the numerical calculation at different gauge parameters 
and use the residual dependence as a check and diagnostic for the 
numerical uncertainty. In this way we verify gauge-parameter independence 
numerically with 0.5 \% accuracy. 
The left panel of Fig.~\ref{fig:gaugedep} shows the residual gauge-parameter 
dependence  of the short-distance coefficient $a_{22}$ for the symmetric 
bulk mass parameters $c_{L_2}=-c_{E_2}=0.5748$. For usual choices for 
the 5D mass parameters the off-shell terms are negligibly small and  
below the numerical accuracy. However, if one of the fermion 
zero modes is IR localized (as would be the case for 
the right-handed top) their effect becomes visible. To illustrate that 
the off-shell terms are required for gauge independence we show in 
the right panel of Fig.~\ref{fig:gaugedep} the gauge-invariant subset 
of diagrams with a $Z_X$ boson that contributes to $a_{ij}$ in the 
custodially protected RS model (see Section~\ref{sec:csRS}).\footnote{
 For diagrams with internal $Z_X$ bosons the numerical accuracy is highest 
as we do not need to subtract zero modes.}
Here $c_{L_i}=0.1$ and $c_{E_j}=-1.2$ are chosen. The plot shows the 
gauge-parameter dependence of the contribution to $a_{ij}$ with 
(upper points) and without (lower points) off shell terms 
normalized to the value for $\xi=1$.

\subsection{Numerical result}

The gauge-boson exchange contribution to $g_\mu-2$ is almost insensitive 
to the structure of the 5D Yukawa matrices and bulk mass parameters and 
can be expressed in terms of known low-energy parameters and the scales 
$T$ and $k$.  To understand this point it is convenient to write the 
gauge-boson contribution $\Delta a^{\rm g}_{ij}$ to the short-distance  
coefficient in the form
\begin{align}
\label{eq:WilsonCoeffReduced}
\Delta a^{\rm g}_{ij}= y^{(5D)}_{ij} \frac{T^3}{k^4}f_{L_i}^{(0)}(1/T) 
g_{E_j}^{(0)}(1/T) \mathcal{A}_{ij}\,,
\end{align}
extracting the Yukawa matrix from the single Higgs insertion in 
Fig.~\ref{fig:diagsall} and the external fermion zero-mode wave-functions. 
We then find numerically that $\mathcal{A}_{ij}\approx \mathcal{A}$ 
varies only mildly with the 5D mass parameters $c_i$, whereas $\Delta 
a^{\rm g}_{ij}$ 
itself is strongly model parameter dependent. However, when 
$\mathcal{A}_{ij}$ does not depend on $ij$, the rotation to the lepton-mass 
eigenbasis simply turns the extracted terms into the diagonal 
lepton-mass matrix, 
hence to very good approximation 
$ \Delta a^{\rm g}_\mu = \sqrt{2}m_\mu/v\times 
\mathcal{A}$, independent of $y^{(5D)}_{ij}$ and bulk masses. For 
the same reason, lepton-flavour violating effects from gauge-boson 
exchange diagrams are strongly 
suppressed.

Thus, the gauge contribution 
to $g_\mu-2$ can the be approximated with accuracy of a few percent by
\begin{equation}
\label{approxGauge}
\Delta a_\mu^{\rm g} \approx 0.25 \cdot 10^{-11} \times \log\frac{k}{T} 
\times\frac{\left(1\,{\rm TeV}\right)^2}{T^2} 
\approx   8.8 \cdot 10^{-11} \times \frac{\left(1\,{\rm TeV}\right)^2}{T^2}
\,.
\end{equation}
The overall scaling factor with $T^{-2}$ is a general feature of 
dimension-six operator effects, while the additional logarithm arises from 
the zero-mode subtracted gauge-boson
propagator. The remaining (implicit) $T$ and $k$ dependence is negligible. 
Note that compared to the four-lepton 
operator contribution (\ref{fourfermioncontribution}) the present one is 
enhanced (rather than suppressed) by the large logarithm $\log(k/T)
\approx 35$. The enhancement arises from the propagation of the 
internal KK fermions. The numerical prefactor 0.25 is however smaller than the 
naive parametric estimate 0.7. Due to the approximate independence of the 
theoretical computation on 
the Yukawa structure and bulk-mass parameters, the muon anomalous 
magnetic moment provides a robust constraint on the KK scale of 
the RS model, similar to the electroweak precision $S$- and $T$-parameters. 

Overall, for a lowest KK excitation of 
mass 1~TeV, the gauge-boson contribution is of order of the present 
experimental and theoretical uncertainty in $g_\mu-2$. It remains, however,  
about a factor 5 below the the present difference between the  
central experimental and theoretical values, though the shift 
$\Delta a_\mu^{\rm g}$ has the right sign to reduce it. Since lowest KK 
excitations with mass 1~TeV are already excluded even by direct 
searches, the anomalous magnetic moment currently provides no 
competitive lower limit on $T$ in the minimal RS model.

\section{5D penguin diagrams -- Higgs exchange}

\begin{figure} 
\begin{center} 
\vskip0.3cm
\includegraphics[width=10cm]{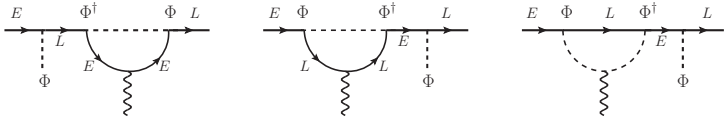}  
\end{center} 
\caption{Higgs-boson exchange diagrams. Non-vanishing contributions 
from these diagrams either require
a wrong-chirality Higgs coupling or the cancellation of the 
external propagator.}  
\label{fig:NonZeroHiggs}  
\end{figure}

The previous discussion ignored contributions from internal Higgs exchange 
diagrams. These diagrams are proportional to a different flavour 
structure, containing three Yukawa coupling factors.\footnote{An odd 
number of Yukawa couplings is required to convert an external SU(2) doublet 
zero mode into an external SU(2) singlet one.} 
There are only three non-vanishing  
one-loop diagrams in the minimal RS model, shown in 
Fig.~\ref{fig:NonZeroHiggs}. Note that the external Higgs field 
necessarily couples to an external fermion line. The coupling to a fermion 
inside the loop leads to dimension-eight operators such as 
$\bar L_i \Phi \sigma^{\mu\nu} E_j F_{\mu\nu} \Phi^\dagger \Phi$.

\begin{figure}
\centering
\includegraphics[width=0.43\textwidth]{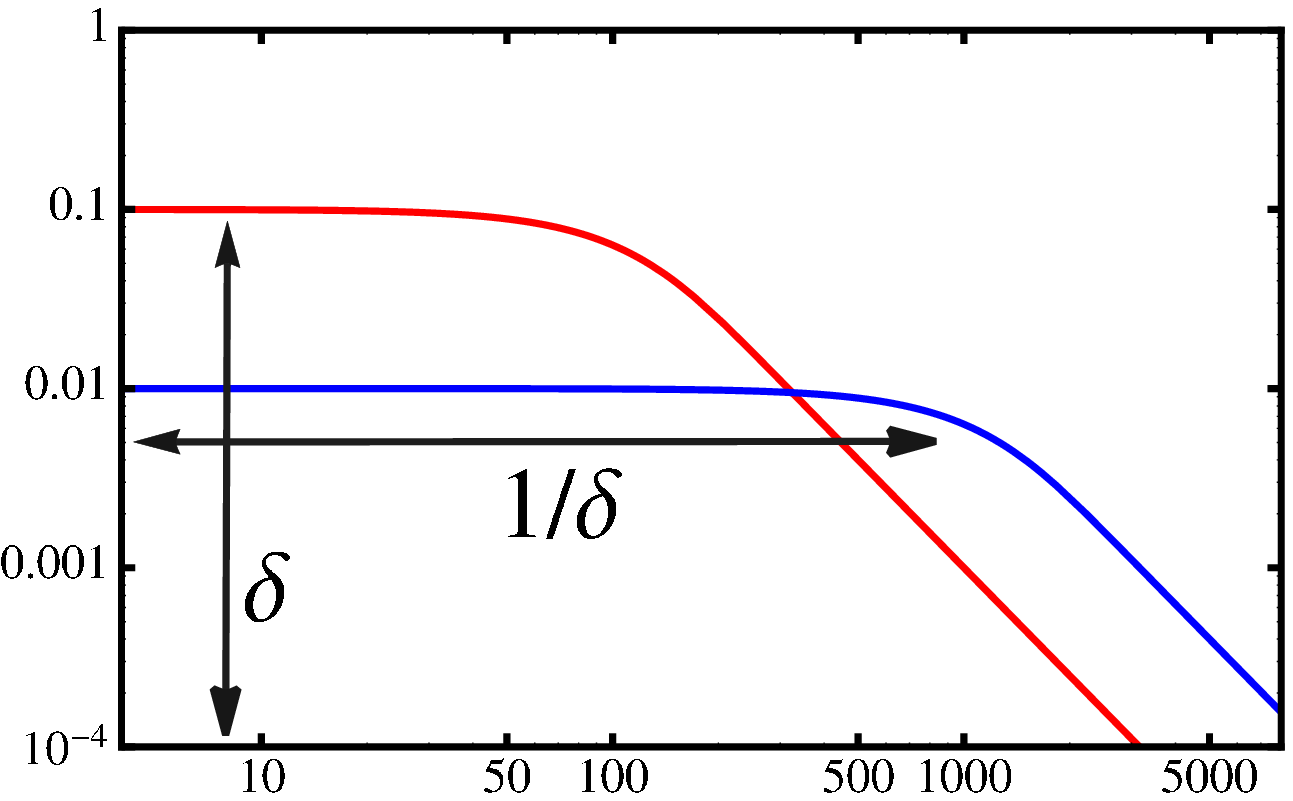}
\hspace*{0.4cm}
\includegraphics[width=0.43\textwidth]{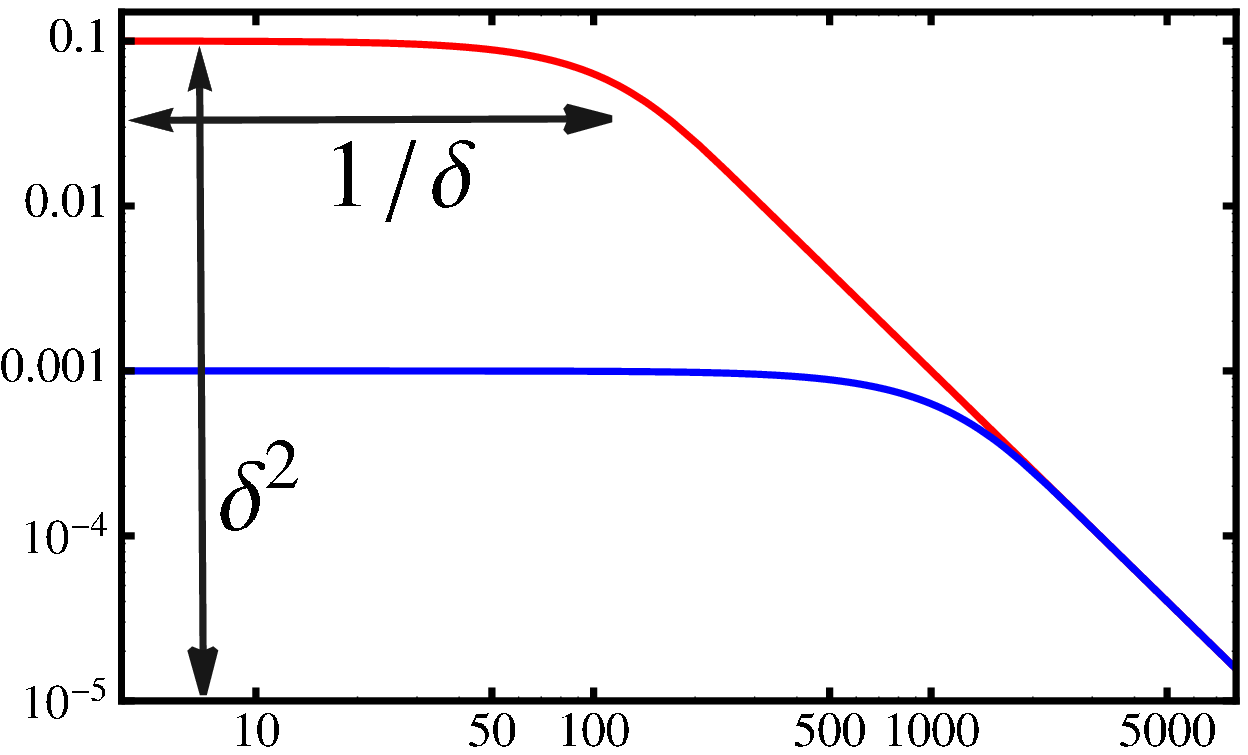}
\vskip0.3cm

\caption{\label{fig:HiggsWidthDependence} 
Qualitative behaviour of the loop momentum integrand (as function of 
$l$ in units of $T$) 
after integration over bulk coordinates and angles leaving only modulus $l$ 
for Higgs (left) and gauge-boson (right) exchange diagrams. The 
blue (dark) curve differs from the red (grey) curve by 
a factor 10 smaller value of $\delta$. }
\end{figure}

Each diagram in Fig.~\ref{fig:NonZeroHiggs} provides two 
distinct contributions to the dipole-operator matching coefficient. 
The first arises from the off-shell vertex function that was discussed 
already in relation with the gauge-boson diagrams.
Its computation is straightforward and can be carried out 
analytically, since due to the brane localization of the Higgs 
field there is only one bulk-coordinate integration from the 
photon vertex. The resulting contribution is quite 
small.\footnote{The smallness 
comes from the external zero-mode propagator. For ``moderate'' choices
of the 5D mass parameters this contribution is usually negligible. This 
is different if one or more of the fermion zero-mode profiles 
are localized towards the IR brane, see Section~\ref{sec:calculation}.}
The second contribution arises from what has been 
discussed in the literature\cite{Csaki:2003sh,Carena:2012fk,Azatov:2009na} 
under the name of ``wrong chirality Higgs couplings'' (WCHCs). 
In the context of radiative transitions, previous references 
to WCHC in the KK picture appear in 
Refs.~\cite{Delaunay:2012cz,Gedalia:2009ws}, where the effect of the lowest
quark KK mode is studied. 
Consider the coupling of the brane-localized Higgs field 
to the SU(2) singlet and doublet leptons, 
 \begin{align}
 \int d^4x \,
\left[(\bar L\Phi) E + \mbox{h.c.}\right]_{z=1/T}=\int d^4x \,
\left[ (\bar L_L \Phi) E_R + (\bar L_R \Phi) E_L + \mbox{h.c.} \right]_{z=1/T} \,,
\end{align}
where the 5D fields are split into their 4D chirality components. 
The second, ``wrong-chirality Higgs coupling'' term on the right-hand 
side is obviously absent in the SM, 
but a priori present in the RS model, since the KK excitations are 
non-chiral. However, the boundary conditions of the right-handed 
SU(2) doublet and left-handed SU(2) singlet require the wrong-chirality 
fields to 
vanish on the IR brane, so the WCHCs vanish for a brane-localized Higgs. 
This expectation turns out to be too naive, since an exactly, 
delta-function localized Higgs cannot be unambiguously 
defined~\cite{Csaki:2003sh}. In the 5D formalism applied to the 
unbroken electroweak theory, an exactly localized Higgs would require 
the evaluation of the discontinuous 5D propagators at the location 
of the discontinuity. To avoid this ambiguity we define the 
RS model with brane-localized Higgs through the limit of a model with 
a Higgs profile with a small width $\delta/T$, where $\delta \ll 1$. 
A possible choice for 
such a regularized profile is 
\begin{align}
\label{higgsprofile}
 \Phi(x,z)= \Phi(x)\,\frac{T}{\delta}\,\Theta(z-(1-\delta)/T)\,.
\end{align}  
For any finite $\delta$ the WCHC are not zero. Moreover, the Higgs 
profile introduces the new scale $T/\delta$
into the problem. It turns out that after integration over all 
bulk coordinates a contribution from the WCHC may survive 
in the limit $\delta\to 0$ that arises from the 
loop momentum region $l\sim T/\delta$.  The 
loop integrand is illustrated in left panel of 
Fig.~\ref{fig:HiggsWidthDependence}, where the 
blue (dark) curve differs from the red (grey) curve by 
a factor 10 smaller value of $\delta$.

The result now depends on the precise meaning 
of ``brane localization'', when the RS model is itself interpreted 
as an effective field theory up to some scale $\Lambda$. ``Exact 
brane localization'' would imply that we take the 
limit $T/\delta\to \infty$ at fixed $\Lambda$. In this case, the 
WCHC contribution vanishes, since the area under the curve 
in the left panel of Fig.~\ref{fig:HiggsWidthDependence} up to 
$l=\Lambda$ is of 
order $\delta\times \Lambda$. However, the RS model with localized 
Higgs can also be interpreted as the idealization of a model with 
a Higgs field that lives parametrically near but not exactly on 
the IR brane. Then, if $T\ll T/\delta\ll \Lambda$, that is, 
when the limit $\Lambda\to \infty$ is taken before $T/\delta\to \infty$, 
we obtain a non-vanishing, model-independent WCHC contribution, since 
the length times height of the plateau in 
Fig.~\ref{fig:HiggsWidthDependence} approaches a finite limit. 
If, however,  $T/\delta\sim \Lambda$, the result depends on the details of 
the Higgs profile.\footnote{A similar non-commutativity of limits 
appears in the RS calculation of Higgs 
production~\cite{Carena:2012fk}, where it was  
discussed in the context of the KK-decomposed theory in the phase of 
broken electroweak theory. }

Focusing on the two limiting cases, we find that 
for $\delta \to 0$ the contribution to the short-distance coefficient 
is independent of the 5D mass parameters and obtain the 
compact expression
\begin{eqnarray}\label{eq:higgscontrib}
\frac{1}{T^2}\,
a_{ij}^{\rm WCHC} = \frac{e}{16\pi^2}\frac{cy^{\rm SM}_{ij}}{T^2} 
\times \frac{[Y Y^\dagger Y]_{ij}}{Y_{ij}} 
\end{eqnarray}
with
\begin{eqnarray}
c \,=\, \left\{\begin{array}{ll} 
-\frac{1}{12} \qquad & \Lambda\to \infty, \mbox{ then }
\delta\to 0\\[0.2cm]
0 & \delta\to 0, \mbox{ then } \Lambda\to \infty
\end{array}\right.
 \end{eqnarray}
and $Y_{ij}=y^{\rm (5D)}_{ij}k$.
Another WCHC Higgs contribution to the electromagnetic dipole transitions 
at order $1/T^2$ comes from the operator 
$h_{ij} \bar L_i\Phi E_j \Phi^\dagger\Phi + \mbox{h.c.}$ 
in (\ref{dimsixlagragian}), since $h_{ij}$ is non-zero only when the 
wrong-chirality Higgs couplings are taken into account. 
The operator is generated at tree-level.  
With the step-function Higgs profile (\ref{higgsprofile}), 
the coefficient function reads (see also Ref.~\cite{Azatov:2009na})
\begin{equation}
\frac{1}{T^2}\,h_{ij} = \frac{y^{\rm SM}_{ij}}{3 T^2} \times 
 \frac{[Y Y^\dagger Y]_{ij}}{Y_{ij}}\,.
\label{hijWCHC}
\end{equation}
When two of the Higgs fields in 
$\bar L_i\Phi E_j \Phi^\dagger\Phi$ are put to their vacuum expectation 
values, this operator modifies the SM Yukawa couplings and leads to 
flavour-changing couplings of the zero-mode fermions to the Higgs boson. 
Inserting this vertex into the Higgs-exchange contribution to the 
electromagnetic dipole transition similar to diagrams (c) and (d) 
of Fig.~\ref{fig:4Ddiags}, we find that the result is suppressed 
relative to (\ref{eq:higgscontrib}) by a factor of [lepton mass]$^2/m_H^2$, 
where $m_H$ is the physical Higgs mass. The additional lepton-mass 
factors arise from the 4D Yukawa coupling at one of the Higgs-fermion vertices 
and the need for a helicity flip in the loop. Thus, the Higgs-exchange 
contribution to the anomalous magnetic moment and to radiative 
lepton-flavour violating transitions from loop momentum $l\sim m_H$ 
is strongly suppressed relative to the contribution 
(\ref{eq:higgscontrib}) that is generated at the KK scale. This 
effect is quite generic and implies that limits on radiative 
lepton-flavour violating decays usually impose much stronger 
constraints on the observability of 
lepton-flavour violating Higgs decays than assumed in 
Ref.~\cite{Harnik:2012pb}.

Before turning back to numerical estimates, we briefly 
comment on potential WCHC contributions to the gauge-boson exchange 
diagrams. While present in principle, we find them vanishing at  
one-loop in the limit $\delta\to 0 $ irrespective
of the order of limits in $\delta$ and $\Lambda$. This can be understood when 
one keeps in mind that in the 5D formalism the WCHC emerge due to the 
discontinuity of the fermion propagator at coincident points.
For a given loop momentum $l$ the leading contribution 
comes from region where the 5D coordinates $x,y$ in the fermion propagator 
$\Delta(l,x,y)$ are within a typical distance of $1/l$. 
A fermion propagator that connects two Higgs vertices is essentially
always near the coincident limit for $\delta \to 0$. For fermion 
propagators that connect a gauge-boson vertex with a Higgs vertex 
the coincidence requirement imposes an additional suppression factor, since 
the gauge boson is not localized near the brane. This is exemplified in 
the right panel of Fig.~\ref{fig:HiggsWidthDependence}, where the 
behaviour of the WCHC terms of diagram B1a/b on $\delta$ is shown. 
The integral over the modulus of $l$ now 
vanishes as $\delta\to 0$, since the height of the plateau scales 
as $\delta^2$.

\section{Combined result}
\label{sec:result}

Adding the Higgs-exchange contributions to the pure gauge-boson 
exchange result (\ref{approxGauge}), we obtain 
\begin{equation}
\label{dares}
\Delta a_\mu \approx   
\left[8.8 + 2.4 \, \langle YY^\dagger \rangle_{\mu} 
\right]
\cdot 10^{-11} \times \frac{\left(1\,{\rm TeV}\right)^2}{T^2}
\,,
\end{equation}
where the dimensionless quantity 
\begin{align} 
\label{eqn_Ytwo}
\langle YY^\dagger \rangle_{\mu}=
\frac{\operatorname{Re}\left[\sum_{lm}[U^\dagger]_{2l}f_{L_l}^{(0)}(1/T) [Y Y^\dagger Y]_{lm} 
g_{E_m}^{(0)}(1/T)V_{m2}\right]}{\sum_{lm}[U^\dagger]_{2l}
f_{L_l}^{(0)}(1/T) Y_{lm} g_{E_m}^{(0)}(1/T)V_{m2}}
\end{align}
parametrises the flavour dependence of the contribution 
from \eqref{eq:higgscontrib} to $g_\mu-2$ 
compared to a term with only a single Yukawa matrix.\footnote{The 
denominator of \eqref{eqn_Ytwo} is 
proportional to the 4D muon Yukawa coupling $y_\mu$.} 
Several comments should be made on this result.

(1) The Higgs-exchange contribution is strongly model-dependent in two ways: 
first, it depends on the entries of the unknown 5D Yukawa matrices. Second, 
it depends on the precise notion of Higgs localization. In particular, 
in the case of exact localization ($\delta \to 0$ first), the term 
proportional to $\langle YY^\dagger \rangle_{\mu}$ is absent altogether. 
The leading Higgs contribution then arises from dimension-eight operators 
and the ``off-shell'' terms. 
A subset of the former was calculated for flavour-violating observables 
in Ref.~\cite{Csaki:2010aj}, where, 
on the other hand, the WCHC terms were not considered.

(2) In the opposite limit ($\Lambda\to \infty$ first) the Higgs 
contribution as given above is the largest contribution to the 
anomalous magnetic moment, if the average Yukawa coupling is somewhat 
larger than one. With experimental and theoretical errors added linearly, the 
measurement of $g_\mu-2$ can then translated into the 
bound 
\begin{align}
\label{eq_HiggsBoundOnY}
-25 < \langle YY^\dagger \rangle_{\mu} \times (1\,\mbox{TeV}/T)^2 < 260\,,
\end{align}
where we required that $a_\mu^{\rm exp}-a_\mu^{\rm RS}$ stays 
compatible with zero at the $3 \sigma$ level.  

(3) When we assume that all entries of the 5D Yukawa matrix are 
of the same order with no cancellations (``anarchic structure''), 
we can constrain the Higgs contribution to $g_\mu-2$ to be 
negligible relative to the gauge-boson one in a model-independent 
way from the non-observation of the $\mu\to e \gamma$ decay. 
Since the gauge contribution to $a_{ij}$ is almost flavour-aligned 
with the mass matrix its effect on FCNCs is suppressed and Higgs 
exchange is the dominant source of the $\mu \to e\gamma$ transition. Assuming 
\begin{align}\label{eq:anarchicCondition}
 \frac{[YY^\dagger Y]_{ij}}{Y_{ij}}\equiv Y_\star^2 
\end{align}
independent of $ij$ (``anarchy''), we find for the branching ratio
\begin{align}
 {\rm Br}(\mu \to e \gamma) = 
6 |c|^2 \,\frac{\alpha_{\rm em}}{4 \pi} 
 \frac{m_e}{m_\mu}\frac{Y_\star^4}{G_F^2 T^4},
\end{align}
where $c=-\frac{1}{12}$ as long as $\Lambda  \gtrsim T/\delta$.
Using the current MEG bound\cite{Adam:2013mnn} 
$ {\rm Br}(\mu \to e \gamma) < 5.7 \cdot 10^{-13}$ 
we obtain 
\begin{align}
 Y_\star \times \frac{1 \; \rm TeV}{T}<0.16\,.
\end{align}
This by itself provides a strong constraint on the size of the 
Yukawa couplings (in combination with the RS scale $T$), which 
corresponds to Kaluza-Klein masses above $15\;\rm TeV $ 
for $Y_\star=1$.

(4) Hence, under the assumption of Yukawa anarchy, we obtain the relation 
\begin{eqnarray}
\Delta a_\mu^{\rm WCHC} &=& 
\sqrt{\frac{m_\mu}{m_e}} \,\frac{G_F m_\mu^2 }{\pi\sqrt{6}e} 
\times \sqrt{\mbox{Br}\,(\mu\to e \gamma)}
\leq 0.6 \cdot 10^{-12}\,,
\end{eqnarray}
independent of the relation of $\delta$ and $\Lambda$
as well as the KK scale $T$. In this case the Yukawa sector and bulk-mass 
independent gauge-boson contribution is by far dominant.

\section{RS model with custodial protection}
\label{sec:csRS}

The minimal RS model is severely constrained by the $T$-parameter, which 
is generated at tree-level with a $\log\frac{k}{T}$ 
enhancement~\cite{Agashe:2003zs}. RS models with custodial protection 
reduce this electroweak precision constraint allowing a lower KK scale. 
At the same time, we expect an enhanced contribution to 
radiative penguin observables due to the larger number of fermion 
and boson states in these models, which can circulate in the loop.

The protection mechanism is based on extending the usual SM hypercharge 
gauge group to a $SU(2)_R\times U(1)_X$ gauge symmetry in 
the bulk~\cite{Agashe:2003zs}. This extended group is then broken 
to $U(1)_Y$ on the UV brane. 
The $U(1)_Y$ boson $B_\mu$ arises as a linear combination 
of $X_\mu$ and $W^3_{R,\mu}$, in analogy to the way the photon is formed from 
$B_\mu$ and $W_{L,\mu}^3$ in the SM.
The orthogonal linear combination is called $Z_X$. 
Only $B_\mu$ has Neumann boundary conditions (BC) on the UV brane;
the remaining bosons ($Z_X$, $W^{1,2}_R$) are endowed with Dirichlet BC. 
This ensures that only the hypercharge boson has a massless mode.  
On the IR brane the vector components of all bosons have the ususal
Neumann BCs; the Higgs mechanism only breaks 
$SU(2)_R \times SU(2)_L$
to the vectorial subgroup and thus large corrections to 
the $T$-parameter are prevented. 
To further forbid large corrections 
to the $Zbb$  vertex from (KK) top quarks an additional discrete 
$\mathds{Z}_2$ symmetry is usually evoked~\cite{Agashe:2006at}. This 
arranges the quark sector in specific gauge multiplets \cite{Carena:2007ua}.
For the lepton sector there is more freedom, 
as the $Z \tau \tau$ vertex does not 
necessarily need an additional protection mechanism.

Here, by extension of the SM, we choose the same multiplet structure for 
leptons and quarks, following Ref.~\cite{Albrecht:2009xr} to which we 
refer for more details. The lepton sector is built from  $\xi_{1L}$ 
(bi-doublet) and  $\xi_{2R}$ (singlet)
as well as $T_3$ and $T_4$ which are singlets under $SU(2)_L$ and 
triplets under $SU(2)_R$: 
\begin{eqnarray}
\xi_{1L}^{il} & = & \left(\begin{array}{cc}
\chi_{L}^{\nu_{i}}\left(-,+\right)_{1} & l_{L}^{\nu_{i}}\left(+,+\right)_{0}\\
\chi_{L}^{l_{i}}\left(-,+\right)_{0} & l_{L}^{l_{i}}\left(+,+\right)_{-1}
\end{array}\right)\nonumber \\
\xi_{2R}^{il} & = & \nu_{R}\left(+,+\right)_{0}\nonumber \\
\xi_{3R}^{il} & = & T_{3R}^{i}\otimes T_{4R}^{i}=\left(\begin{array}{c}
\tilde{\lambda}_{R}^{_{i}}\left(-,+\right)_{1}\\
\tilde{N}_{R}^{_{i}}\left(-,+\right)_{0}\\
\tilde{L}_{R}^{_{i}}\left(-,+\right)_{-1}
\end{array}\right)\otimes\left(\begin{array}{c}
\lambda_{R}^{_{i}}\left(-,+\right)_{1}\\
N_{R}^{_{i}}\left(-,+\right)_{0}\\
E_{R}^{_{i}}\left(+,+\right)_{-1}
\end{array}\right).
\end{eqnarray}
The subscript on the different fermion fields indicates the electric
charge $Q$. The signs in parentheses refer to 
 the BC on the UV (left) and IR brane (right); 
a ``+'' corresponds to Neumann and a ``$-$'' to Dirichlet BC.
This extended fermion sector introduces new topologies for 
the one loop-diagrams that are not present in the minimal model. 
The total number of diagrams more than doubles, but only a small subset 
leads to new integral structures.
The basic strategy for the calculation remains the same. As expected, 
the extended gauge and fermion sector leads to an enhanced gauge-boson 
contribution of to $g_\mu-2$ \cite{inprep}:
\begin{align}
 \Delta a_\mu\approx 27.2 \cdot 10^{-11} 
\times\left(\frac{1\,\mbox{TeV}}{T}\right)^2
\end{align}
 --- more than a factor of three larger than for the minimal 
model.\footnote{This enhancement would be absent for the simplest lepton 
multiplet structure, where only the SM singlet is promoted to an 
$SU(2)_R$ doublet.} 
The dependence on $T$ and the relative insensitivity of the 
gauge-boson exchange contributions to 5D masses and Yukawa
parameters are general features of RS models with a localized 
Higgs interactions and do not depend on the precise details of the model. 

The Higgs contributions are, in general, model-dependent just as 
in the minimal case. The main new aspect is the presence of two 
Yukawa matrices instead of only one. One Yukawa matrix 
governs the interaction between the bi-doublet and the singlet, while the 
other one couples the bi-doublet and triplet. The presence of two 
Yukawa couplings allows for cancellations which make general statements
on the size of the total contributions even more difficult.
In particular, if both Yukawas are equal, the dominant contributions 
cancel, and the Higgs contribution to the dipole operator 
coefficients becomes negligible.   

\section{Summary}
   
We presented a complete computation of leptonic radiative penguin diagrams
in the minimal and custodially protected Randall-Sundrum model. 
To this end we performed a matching calculation onto  
SU(3)$\times$SU(2)$\times$U(1)-invariant dimension-six operators 
in the unbroken electroweak theory by integrating out the bulk of 
the warped space-time in a manifestly 5D framework. 

Penguin diagrams with gauge bosons in the loop turn out to be  
technically challenging as the calculation involves up to three bulk 
coordinate and one 4D loop-momentum integrations. Moreover, while finite, 
the calculation is only consistent 
when carried out with a regulator. 
Without a regularization the sum of short-distance 5D loops and 4D loops with 
insertions of higher dimensional operators is scheme dependent and cannot be 
associated with a physical observable. 

Surprisingly, the gauge-boson mediated penguin amplitudes  
are quite insensitive to the 5D Yukawa structure and bulk masses. 
This allowed us to 
derive a model-independent result (in the sense of being dependent 
essentially only on the scale of the model, but not on its other 
parameters) for their contribution to the muon 
anomalous magnetic moment.

On the other hand, penguin diagrams that are generated by an internal
Higgs boson exchange turn out to be straightforward to compute analytically
in dimensional regularization. The main subtlety arises from the sensitivity 
to the precise implementation of the IR-brane localization of the Higgs.
Once this is specified the result is unambiguous, but depends strongly 
on the parameters of the 5D Lagrangian.

The dimension-six Lagrangian can be utilized to study the consequences of 
RS models in typical penguin-induced processes like flavour-changing 
radiative lepton $\ell_i \to \ell_j\gamma$ and the  
magnetic (and electric) dipole moments.  
We find that the contribution to $g_\mu-2$ is enhanced by 
$\log\frac{k}{T}\approx 
35$ compared to the naive dimensional analysis. In the minimal RS set-up the 
model-parameter  independent gauge-boson contribution increases the 
value of  $a_\mu$.  The effect has the right sign towards the 
present experimental measurement, but the shift
\begin{align}
 \Delta a_\mu\approx 8.8 \cdot 10^{-11} 
\times\left(\frac{1\,\mbox{TeV}}{T}\right)^2
\end{align}
 is too small to resolve the discrepancy of measured value and  
theory prediction for viable KK masses.\footnote{Recall that the 
lowest KK excitation's mass is around $2.5 T$.} 
In RS models with an extended bulk gauge and fermion sector like the
custodially protected model we generally find larger contributions.
In the particular set-up studied in Section~\ref{sec:csRS} the shift is 
more than a factor of three larger than in the minimal model. 

Higgs contributions to the anomalous magnetic moment depend strongly 
on the size of Yukawa couplings, but are negligible in anarchic 
models due to the constraints from lepton-flavour violating 
decays. These in turn imply strong restrictions on
the Yukawa structure and KK scale, which deserve a more detailed 
study.

\section*{Acknowledgments}

\noindent 
This work has
been supported by the DFG SFB/TR~9
``Com\-pu\-ter\-gest\"utzte Theoretische Teil\-chen\-physik,'' the Gottfried
Wilhelm Leibniz programme of the Deutsche Forschungsgemeinschaft (DFG),
and the DFG cluster of excellence ``Origin and Structure of the
Universe.'' The work of JR is supported by STFC UK.


\bibliographystyle{ws-ijmpa}

\end{document}